\documentclass[12pt]{article}

\usepackage{amsmath, amssymb, amsthm} 
\usepackage{graphicx} 
\usepackage{cite} 
\usepackage{hyperref} 
\usepackage{geometry} 
\usepackage{graphicx} 

\usepackage{xcolor} 
\usepackage{algorithm}
\usepackage{algpseudocode}
\usepackage{algorithm}

\usepackage{subcaption} 

\usepackage{array} 

\usepackage{placeins} 

\title{Identifying Core-Periphery Structures in Networks via Artificial Ants}
\author{Imran Ansari$^{1}$,  Qazi J Azhad$^{1}$, Niteesh Sahni$^{1}$ \\ 
\small $^{1}$Department of Mathematics, Shiv Nadar Institution of Eminence, Delhi-NCR, India \\
\small \texttt{Email: ia717@snu.edu.in, qazi.jamal@snu.edu.in, niteesh.sahni@snu.edu.in} 
}
\date{} 

\begin{document}
\maketitle

\begin{abstract}

Core-periphery structure represents a meso-scale structure in networks, characterized by a dense interconnection of core nodes and sparse connections among peripheral nodes. In this paper, we introduce an innovative approach for detecting core-periphery structure, leveraging Artificial Ants. Core-periphery structures play a crucial role in elucidating network organization across various domains. The proposed approach, inspired by the foraging behavior of ants, employs artificial pheromone trails to iteratively construct and refine solutions, thereby eliminating the need for arbitrary partitions that often constrain traditional methods. Our method is applied to a diverse selection of real-world networks—including historical, literary, linguistic, sports, and animal social networks—highlighting its adaptability and robustness. We systematically compare the performance of our approach against established core-periphery detection techniques, emphasizing differences in node classification between the core and periphery. Experimental results show that our method achieves superior flexibility and precision, offering marked improvements in the accuracy of core-periphery structure detection.

\end{abstract}

\section{Introduction}


In network science, meso-scale structures in networks have garnered significant attention in recent years, as detecting these intermediate-scale patterns algorithmically enables the discovery of network characteristics that are not readily observable at the local level of nodes and edges or through global summary statistics\cite{Borgatti1999,kojaku2017finding,Rombach2017,Boyd2010}. Among the various types of meso-scale structures that emerge in networks, core-periphery structures have received particular focus due to their distinct organization and their relevance across diverse domains. Core-periphery structures consist of densely connected core nodes that contribute to the network's cohesiveness and are linked to peripheral nodes, which are sparsely interconnected among themselves\cite{Borgatti1999, Rossa2013,holme2005core}. This configuration reflects a hierarchical and functional organization within the network. Over the past few decades, numerous applications of the core-periphery structure have been proposed in various domains, including social networks \cite{Borgatti1999, Boyd2010, Rossa2013}, protein-protein interaction (PPI) networks \cite{Rossa2013, yang2014overlapping}, financial networks \cite{in2020formation, craig2014interbank}, transportation networks \cite{Rombach2017, Rossa2013}, and neural networks \cite{Rossa2013, tunc2015}. Various methods have been developed to detect and analyze core-periphery structures, underscoring their importance in understanding network dynamics and information flow across different domains \cite{Rossa2013, Boyd2010, Rombach2017, kojaku2017finding, Borgatti1999}.

Optimization problems are of great significance in both industrial and scientific domains.  Ant Colony Optimization(ACO), introduced in the early 1990s, is a metaheuristic designed to solve complex combinatorial optimization problems \cite{dorigo2003ant, dorigo1999ant, dorigo2006ant}. The ACO algorithm is inspired by the foraging behavior of real ants, which communicate using pheromones. In a similar fashion, ACO uses artificial pheromone trails to enable indirect communication among a colony of artificial ants\cite{dorigo2004}. These pheromone trails represent distributed numerical information that the ants use to construct and refine solutions to the problem as the algorithm progresses, incorporating their past search experiences. ACO has been successfully applied to a wide range of domains, including the Traveling Salesman Problem (TSP) \cite{dorigo2003ant, dorigo1996ant}, Quadratic Assignment Problem \cite{maniezzo1999exact}, Scheduling Problems \cite{den2000ant}, Vehicle Routing Problems \cite{gambardellamultiple}, Graph Coloring \cite{costa1997ants}, Shortest Supersequence Problem \cite{michel1998island}, Sequential Ordering \cite{gambardella2000ant}, Data Mining \cite{parpinelli2002data}, Maximum Clique Problem \cite{bui2004finding}, Edge-Disjoint Paths Problem \cite{blesa2004ant}, Multi-Objective Problems \cite{lopez2004design}, Dynamic and Stochastic Problems \cite{bianchi2002ant}, and even in the domain of Music \cite{gueret2004ants}. The first application of ACO was on the TSP through the original Ant System (AS) \cite{dorigo1991positive, dorigo1996ant}, a paradigmatic NP-hard combinatorial optimization challenge that has been extensively studied \cite{cook2011traveling, johnson1997traveling}. TSP continues to serve as a benchmark for evaluating new algorithmic ideas and variations. Although the AS algorithm produced promising results for TSP, it was later found to be less effective compared to state-of-the-art algorithms for TSP and other combinatorial optimization problems, leading to the development of various extensions and enhancements over the years\cite{dorigo2004}.


Numerous successful ACO variants have been proposed, including Elitist AS (EAS) \cite{dorigo1996ant}, Rank-based AS (RAS) \cite{bullnheimer1999new}, MAX–MIN Ant System (MMAS) \cite{stutzle2000max}, Ant Colony System (ACS) \cite{dorigo1997ant}, Hyper-Cube Framework (HCF) \cite{blum2004hyper}, Ant-Q \cite{fuellerer2009ant}, ANTS \cite{maniezzo1999exact}, Best-Worst AS \cite{cordon2002analysis}, Population-based ACO \cite{guntsch2002population}, Beam-ACO \cite{blum2004theoretical}, and many more. The ACO metaheuristic provides a versatile framework for both existing applications and the development of algorithmic variants \cite{dorigo2003ant, dorigo1999ant}. ACO algorithms use a probabilistic construction heuristic based on artificial pheromone trails and possibly available heuristic information from the problem's input data\cite{dorigo2004}. This allows ACO to be seen as an extension of traditional construction heuristics, which are often available for many combinatorial optimization problems. A key difference, however, is the adaptive nature of pheromone trails, which evolve during the execution of the algorithm, reflecting the accumulated search experience.

 The concept of core-periphery structure in networks, as initially formulated by Borgatti et al.~\cite{Borgatti1999}, is based on the intuition that core nodes are highly interconnected, there are connections between core and periphery nodes, and no connections exist between periphery nodes. Let \( G \) represent an undirected, unweighted network containing \( N \) vertices and \( M \) edges, with no self-loops or multiple edges. Define \( A = (a_{ij}) \) as the adjacency matrix of \( G \), where \( a_{ij} = 1 \) if nodes \( i \) and \( j \) are connected, and \( a_{ij} = 0 \) otherwise. A network \( G \) is recognized as a core-periphery network if a subset of core nodes \( K \subset N \) and a subset of periphery nodes \( P \subset N \setminus K \) can be identified, such that:
\begin{itemize}
    \item[i.] For all \( i, j \in K \): \( a_{ij} = 1 \).
    \item[ii.] For all \( i, j \in P \): \( a_{ij} = 0 \).
    \item[iii.] For each \( i \in K \), there exists at least one \( j \in P \) with \( a_{ij} = 1 \); similarly, for each \( j \in P \), there exists at least one \( i \in K \) with \( a_{ij} = 1 \).
\end{itemize}

According to Borgatti et al.~\cite{Borgatti1999}, a label \( c_i \) is assigned to each node \( i \) such that \( c_i = 1 \) if the node \( i \) is in the core set \( K \), and \( c_i = 0 \) if it belongs to the periphery \( P \). The sets \( K \) and \( P \) are determined by maximizing the function 
\[
\rho(c_1,\ldots,c_N) = \sum_{i=1}^{N} \sum_{j=1}^{N} a_{ij} c_{ij}
\]
over all values \( c_1, \ldots, c_N \), where \( c_{ij} = 1 \) if either node \( i \) or \( j \) is in \( K \), and \( c_{ij} = 0 \) if both nodes are in \( P \). Borgatti et al.~\cite{Borgatti1999} also introduce the idea of assigning a "\textit{coreness}" value to each node by defining a parameter \( c_i \in [0,1] \) for each vertex \( i \), computed by maximizing the expression 
\[
\rho(c_1, \ldots, c_N) = \sum_{i=1}^{N} \sum_{j=1}^{N} a_{ij} c_i c_j.
\]

 Inspired by the continuous version of Borgetti's core-periphery structure detection model, Rombach et al.~\cite{Rombach2017} developed an extension where the core quality is defined via a transition function that is optimized to obtain a core score for each node. In Refs\cite{Rossa2013}, Rossa et al. introduce a method for detecting the core-periphery structure through an iterative algorithm that yields the core-periphery profile by elaborating on the behavior of a random walker. Boyd et al.~\cite{Boyd2010} aim to find a MINRES vector \( \mathbf{c} \) such that the adjacency matrix \( \mathbf{A} \) is approximated by \( \mathbf{c}\mathbf{c}^T \). This approximation minimizes the sum of squared differences off the diagonal. Specifically, it seeks to find a vector \( \mathbf{c} \) that minimizes

\[
\sum_{i} \sum_{j \neq i} [a_{ij} - c_i c_j]^2.
\]

By taking the partial derivative with respect to each element of \( \mathbf{c} \), we obtain

\[
c_i = \frac{\sum_{j \neq i} a_{ij} c_j}{\sum_{j \neq i} c_j^2},
\]

which leads to an iterative process for computing the MINRES vector.

In the present work, we propose a model for detecting the core-periphery structure in networks using  ACO. We evaluate the proposed method through numerical experiments on real-world data from various domains, including historical networks, literary networks, linguistic networks, sports networks, animal social networks, and political networks. 
We validate the correctness of the core scores learned by comparing them with well-established core score estimation algorithms. The results indicate that the proposed method effectively estimates core scores of the vertices and performs on par with existing methods.

The remainder of this paper is organized as follows. In Section 2, the authors provided a description and methodology of proposed method for detecting core-periphery structures. In Section 3, the proposed method is applied to various datasets and results are discussed with detailed interpretation. The description of each dataset is also provided along with results. Finally, conclusion of this study is provided.

\section{Proposed Method}

The objective of our approach is to detect core-periphery structures within a network by leveraging the  ACO framework, focusing on maximizing the weight of connections between core nodes while minimizing less significant links.

Let \( G = (V, E) \) be a weighted, simple, and undirected graph, where \( V \) denotes the set of vertices, and \( E \) represents the set of edges. Each edge \( (i, j) \in E \) is assigned a weight \( w_{ij} \), representing the strength of the connection between vertices \( i \) and \( j \). If \( w_{ij} = 0 \) (i.e., there is no weight between vertices \( i \) and \( j \)), then \( w_{ij} \) is set to a very small positive value.

 ACO algorithms are well-suited to network analysis due to their iterative solution refinement process, which mimics the behavior of ants searching for optimal paths. For our analysis, we employ the original Ant System (AS), which iteratively updates pheromone values on edges, allowing us to identify high-weight, core edges in the network. The pheromone update mechanism, a critical part of AS, reinforces the edges that form part of high-quality solutions (i.e., stronger core-periphery structures). The update formula for pheromone levels on edge \( (i, j) \) is as follows:

\begin{equation}
\tau_{ij} \leftarrow (1 - \rho) \cdot \tau_{ij} + \sum_{k=1}^{m} \Delta \tau_{ij}^k
\label{eq:tau_update}
\end{equation}

where \( \rho \in (0,1] \) is the pheromone evaporation rate, preventing the system from overcommitting to early solutions, and \( m \) is the number of ants. 

\(
\Delta \tau_{ij}^k 
\) is the pheromone deposited on edge \( (i, j) \) by ant \( k \), defined as:

\begin{equation}
\Delta \tau_{ij}^k =
\begin{cases}
\frac{1}{L_k}, & \text{if ant } k \text{ uses edge } (i,j), \\
0, & \text{otherwise}
\end{cases}
\label{eq:delta_tau}
\end{equation}

where \( L_k \) represents the total accumulated weight of the path constructed by ant \( k \).




The transition probability \( p_{ij}^k \) of ant \( k \), currently located at vertex \( i \), moving from vertex \( i \) to \( j \) is given by:

\begin{equation}
p_{ij}^k =
\begin{cases}
\frac{(\tau_{ij})^{\alpha} \cdot (\eta_{ij})^{\beta}}{\sum\limits_{l \in N_i^k} (\tau_{il})^{\alpha} \cdot (\eta_{il})^{\beta}} & \text{if } j \in N_i^k, \\
0 & \text{otherwise}
\end{cases}
\label{eq:p_ij}
\end{equation}

where \( N_i^k \) denotes the set of feasible neighbors of ant \( k \) when at vertex \( i \). The parameters \( \alpha \) and \( \beta \) control the relative importance of the pheromone and heuristic information \( \eta_{ij} \), where \( \eta_{ij} = w_{ij} \), representing the weight between vertices \( i \) and \( j \). \\
The pseudo-code for the solution construction procedure is presented as Algorithm \ref{alg:core_periphery}. In Ref.~\cite{dorigo2004} \textit{(Theorem 4.1)}, it is demonstrated that the probability of the algorithm finding an optimal solution can be made arbitrarily close to 1 by selecting a sufficiently large number of iterations, \( n \).





\begin{algorithm}
\caption{Pseudo-code for the solution construction procedure}
\label{alg:core_periphery}
\begin{algorithmic}
\State Initialize pheromone levels $\tau_{ij}$ for all edges $(i, j) \in E$
\State Initialize parameters: $m$ (number of ants), $\rho$ (pheromone evaporation rate), $\alpha$, $\beta$ (influence of pheromone and heuristic information), and $\text{max\_iter}$ (maximum iterations)
\State Initialize graph $G = (V, E)$ with weights $w_{ij}$ on each edge $(i, j) \in E$
\State Set $iteration \gets 0$
\While{$iteration < \text{max\_iter}$}
    \For{each ant $k$}
        \State Initialize \texttt{ant[k].tour} as empty
        \State Initialize \texttt{ant[k].visited} as \texttt{false} for all vertices in $V$
        
        \Comment{Step 1: Ant selects a random initial node}
        \State Select random initial node $r$ for \texttt{ant[k]}
        \State \texttt{ant[k].tour.append}($r$)
        \State \texttt{ant[k].visited}[$r$] = \texttt{true}
        
        \Comment{Step 2: Construct the tour}
        \State $step \gets 1$
        \While{$step < n$  (where $n$ is the total number of nodes)}
            \For{each possible node $j$ that \texttt{ant[k]} can visit}
                \State Compute transition probability $p_{ij}^k$ using pheromone $\tau_{ij}$ and heuristic information $\eta_{ij}$:
                \[
                p_{ij}^k = \frac{(\tau_{ij})^{\alpha} \cdot (\eta_{ij})^{\beta}}{\sum_{l \in N_i^k} (\tau_{il})^{\alpha} \cdot (\eta_{il})^{\beta}}
                \]
            \EndFor
            \State Select next node $j$ based on $p_{ij}^k$ (probabilistic selection)
            \State \texttt{ant[k].tour.append}($j$)
            \State \texttt{ant[k].visited}[$j$] = \texttt{true}
            \State $step \gets step + 1$
        \EndWhile
    \EndFor

    \Comment{Step 3: Update pheromone levels}
    \For{each edge $(i, j) \in E$}
        \For{each ant $k$}
            \If{\texttt{ant[k]} uses edge $(i,j)$}
                \State $\Delta \tau_{ij}^k = \frac{1}{L_k}$ \Comment{where $L_k$ is the total weight of the path constructed by ant $k$}
            \EndIf
        \EndFor
        \State Update pheromone: $\tau_{ij} \gets (1 - \rho) \cdot \tau_{ij} + \sum_{k=1}^{m} \Delta \tau_{ij}^k$
    \EndFor
    \State Increment $iteration \gets iteration + 1$
\EndWhile

\State \textbf{Return} the solution (final pheromone levels)
\end{algorithmic}
\end{algorithm}


We define the cohesiveness measure \( \Psi_S \) for a subgraph \( S \subseteq G \) as:

\begin{equation}
\Psi_S = \frac{\sum_{i \in S, j \in S} (\tau_{ij})^{\alpha} (\eta_{ij})^{\beta}}{\sum_{l \in S, i \in V(G)} (\tau_{il})^{\alpha} (\eta_{il})^{\beta}}
\label{eq:psi_S}
\end{equation}

This measure \( \Psi_S \) reflects the likelihood that an ant, currently positioned at any vertex in \( S \), remains within \( S \) in the next step, thereby serving as an indicator of the subset's cohesiveness within the network.

The determination of the \textit{coreness} for each vertex in the network is carried out through an iterative procedure. The process begins by selecting the vertex with the smallest weighted degree and defining the set containing this vertex as \( S_1 \). Without loss of generality, let \( S_1 = \{1\} \), and set the cohesiveness measure \( \Psi_1 := \Psi_{S_1} = 0 \).

In the subsequent step, we consider the subsets \( S_2^{(j)} := S_1 \cup \{j\} \) for all \( 2 \leq j \leq N \) and compute the cohesiveness measure \( \Psi_{S_2^{(j)}} \) for each subset. Let \( \Psi_{S_2^{(k)}} \) represent the minimum cohesiveness measure among all the subsets. We then update the set to \( S_2 := S_2^{(k)} \) and assign the coreness value \( \Psi_2 := \Psi_{S_2^{(k)}} = \Psi_{S_2} \), which corresponds to the \textit{coreness} of vertex \( k \). Notably, \( S_2 \) now consists of two vertices with the lowest transition probability, ensuring the condition \( \Psi_1 \leq \Psi_2 \).

This procedure is iterated to construct \( S_3 \) from \( S_2 \), compute \( \Psi_3 \), the coreness of the third vertex, and maintain the inequality \( \Psi_1 \leq \Psi_2 \leq \Psi_3 \). The process continues in this manner until all vertices are included in the set. Ultimately, we obtain the sequence \( \Psi_1 \leq \Psi_2 \leq \cdots \leq \Psi_N \). The pseudo-code for computing the \textit{coreness} value is provided in Algorithm~\ref{alg:coreness}.



\begin{algorithm}
\caption{Pseudo-code for Determining Vertex \textit{Coreness} in a Network}
\label{alg:coreness}
\begin{algorithmic}
\State \textbf{Input:} Graph $G = (V, E)$ with parameters $\tau_{ij}$ and $\eta_{ij}$ for edges
\State \textbf{Output:} \textit{Coreness} values $\Psi$ for each vertex

\State Initialize \( S_1 \gets \{1\} \) \Comment{Start with vertex of least weighted degree}
\State Set \( \Psi_1 \equiv \Psi_{S_1} = 0 \) \Comment{Initialize cohesiveness measure}

\While{there are vertices not included in \( S \)}
    \For{each vertex \( j \notin S \)}
        \State Define \( S^{(j)} = S \cup \{j\} \)
        \State Compute \( \Psi_{S^{(j)}} \) using Eq.\eqref{eq:psi_S}
    \EndFor
    \State Select vertex \( k \) such that \( \Psi_{S^{(k)}} \) is minimized
    \State Update \( S \gets S \cup \{k\} \) and set \( \Psi \) value for \( k \)
\EndWhile

\State \textbf{Return:} Coreness values \( \Psi \) for all vertices in \( V \)
\end{algorithmic}
\end{algorithm}


\textbf{\textit{Remark 1}}: In an unweighted, simple, and undirected graph \( G \), where \( w_{ij} = 1 \) for adjacent vertices and \( w_{ij} = 0 \) otherwise, the heuristic information \( \eta_{ij} = 1 \). The cohesiveness measure simplifies to:

\begin{equation}
\Psi^{u}_S = \frac{\sum_{i \in S, j \in S} \left( \tau_{ij} \right)^{\alpha}}{\sum_{l \in S, i \in V(G)} \left( \tau_{il} \right)^{\alpha}}
\label{eq:phi_S_unweighted}
\end{equation}

\textbf{\textit{Remark 2}}: In a weighted graph where each edge \( (i, j) \in E \) is assigned a distance \( d_{ij} \), the heuristic information \( \eta_{ij} = \frac{1}{d_{ij}} \). The cohesiveness measure becomes:

\begin{equation}
\Psi^{d}_S = \frac{\sum_{i \in S, j \in S} \left( \tau_{ij} \right)^{\alpha} \left( \frac{1}{d_{ij}} \right)^{\beta}}{\sum_{l \in S, i \in V(G)} \left( \tau_{il} \right)^{\alpha} \left( \frac{1}{d_{il}} \right)^{\beta}}
\label{eq:psi_S_dist}
\end{equation}\\




\section{Results \& Discussion}
The datasets utilized in this study cover a wide array of network types. These include Zachary's Karate Club \cite{zachary1977information}, Florentine Families \cite{breiger1986cumulated}, Davis Southern Women \cite{davis1941deep}, Krackhardt Kite \cite{krackhardt1990assessing}, Les Misérables \cite{knuth1993stanford}, Word Adjacencies \cite{newman2006finding}, American College Football \cite{lusseau2003bottlenose}, Dolphins \cite{lusseau2003bottlenose}, and Books About US Politics (compiled by V. Krebs, available at \url{http://www.orgnet.com}).\\
To assess the efficiency of our proposed method, we compare it with well-established methods such as Roosa et al.\cite{Rossa2013}, Rombach et al.\cite{Rombach2017}, and Boyd\cite{Boyd2010} methods, we followed the approach outlined below. First, we permuted the rows and columns of the ground truth adjacency matrices in decreasing order of the core scores output by different methods and normalized the matrices so that their entries lie between 0 and 1. We denote the normalized permuted matrix as \( \Phi_0 \). We compare the ground truth adjacency matrix with the adjacency matrix of the ideal core-periphery structure \cite{Borgatti1999}. In the ideal core-periphery structure, core nodes are adjacent to other core nodes, core nodes are adjacent to periphery nodes, but periphery nodes are not adjacent to each other. The adjacency matrix corresponding to the idealized core-periphery structure is defined as:

\begin{equation}
\Phi_{\text{ideal}} =
\begin{pmatrix}
\mathbf{1}_{k \times k} & \mathbf{1}_{k \times (N-k)} \\
\mathbf{1}_{(N-k) \times k} & \mathbf{0}_{(N-k) \times (N-k)}
\end{pmatrix}
\label{eq:phi_ideal}
\end{equation}

where \( \mathbf{1}_{m \times n} \) and \( \mathbf{0}_{m \times n} \) denote matrices filled with ones and zeros, respectively. In this model, \( k \) represents the number of core nodes, and we set \( k \) to \( N/4 \) for all the datasets.

For comparison, we computed the Frobenius norm difference between the normalized permuted matrix \( \Phi_0 \) and the idealized core-periphery matrix \( \Phi_{\text{ideal}} \), which is defined as:

\begin{equation}
\|\Phi_0 - \Phi_{\text{ideal}}\|_F = \sqrt{\sum_{i=1}^{m} \sum_{j=1}^{n} |\Phi_{0_{ij}} - \Phi_{\text{ideal}_{ij}}|^2}
\label{eq:frobenius_norm}
\end{equation}

where \( \Phi_{0_{ij}} \) and \( \Phi_{\text{ideal}_{ij}} \) are the \((i,j)\)th elements of \( \Phi_0 \) and \( \Phi_{\text{ideal}} \), respectively.\\

To further evaluate their performance, we calculate the cosine similarity between the \textit{coreness} values obtained by the proposed method and those from the well-established methods. 
The cosine similarity \( S_c(\mathbf{X}, \mathbf{Y}) \) between two non-zero vectors \( \mathbf{X} \) and \( \mathbf{Y} \) is defined as:

\begin{equation}
S_c(\mathbf{X}, \mathbf{Y}) := \frac{\mathbf{X}^T \mathbf{Y}}{\|\mathbf{X}\|_2 \|\mathbf{Y}\|_2}
\label{eq:sc}
\end{equation}


Before presenting a comprehensive comparison, we first focus on optimizing the parameters $\alpha$, $\beta$, and $\rho$, as these play a critical role in the performance of the algorithms.

\begin{figure}[htbp] 
    \centering
    \begin{subfigure}{0.45\textwidth}
        \centering
        \includegraphics[width=\textwidth]{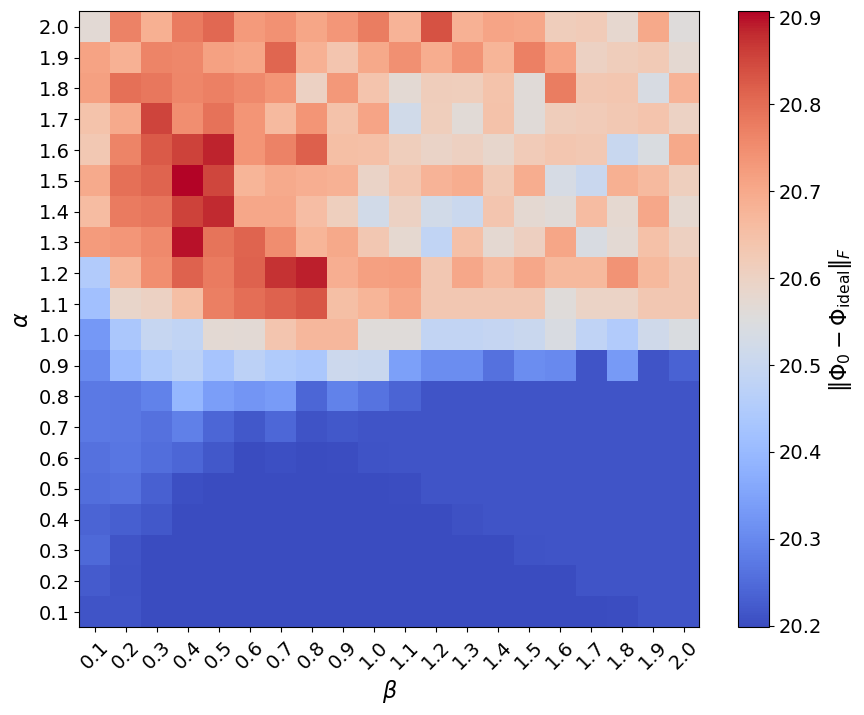}
        \caption{$\rho=0.1$}
        \label{fig:karate_rho_0_1}
    \end{subfigure}
    \hfill
    \begin{subfigure}{0.45\textwidth}
        \centering
        \includegraphics[width=\textwidth]{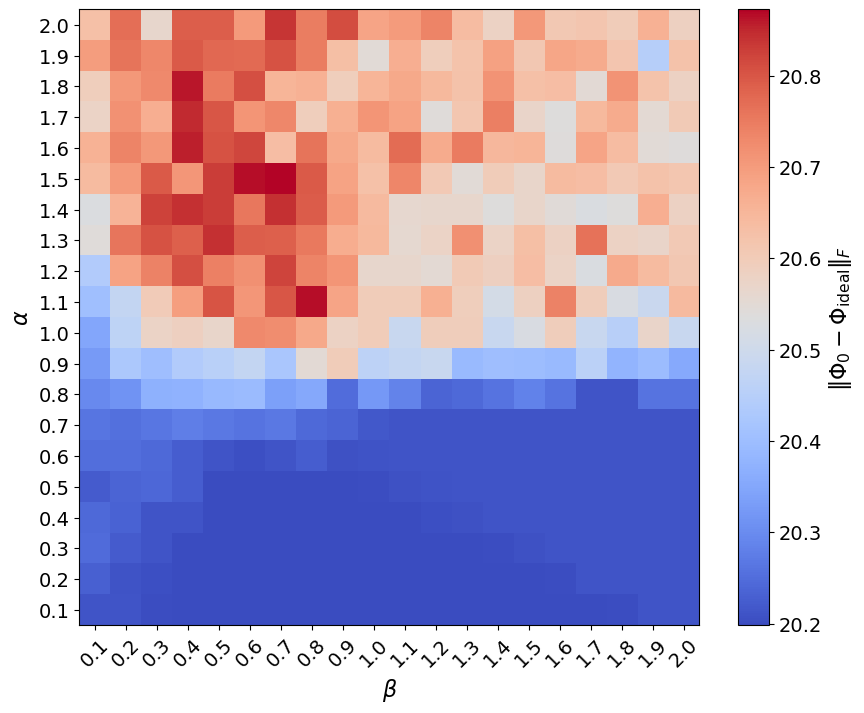}
        \caption{$\rho=0.3$}
        \label{fig:karate_rho_0_3}
    \end{subfigure}
    
    \vspace{0.5cm} 
    
    \begin{subfigure}{0.45\textwidth}
        \centering
        \includegraphics[width=\textwidth]{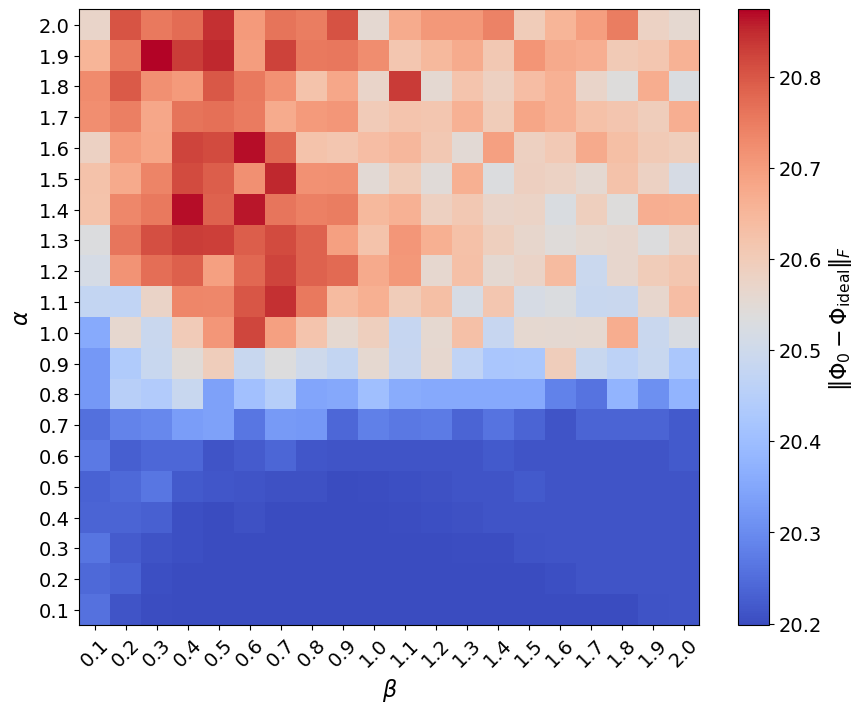}
        \caption{$\rho=0.5$}
        \label{fig:karate_rho_0_5}
    \end{subfigure}
    \hfill
    \begin{subfigure}{0.45\textwidth}
        \centering
        \includegraphics[width=\textwidth]{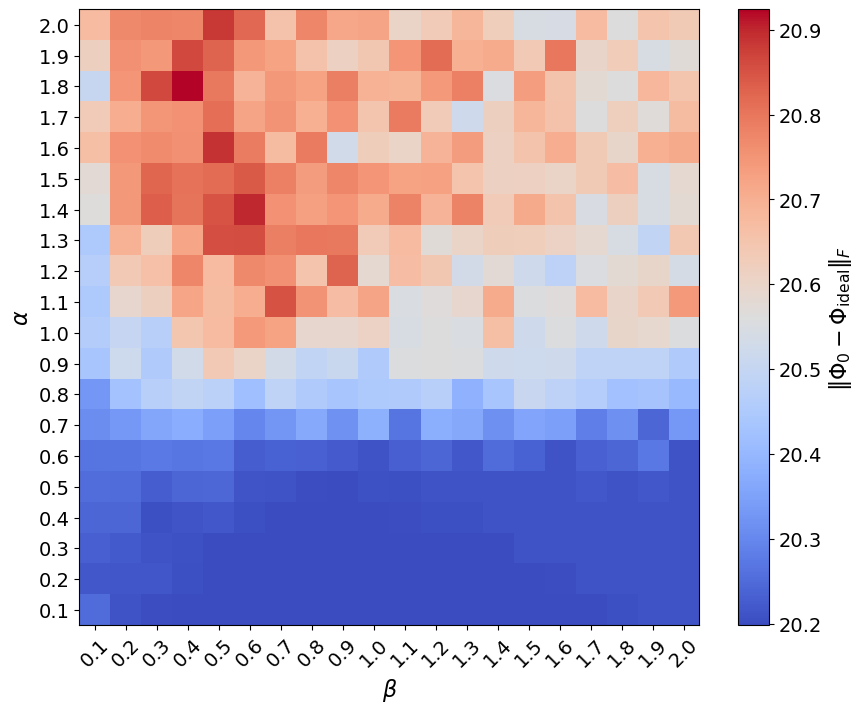}
        \caption{$\rho=0.7$}
        \label{fig:karate_rho_0_7}
    \end{subfigure}
    
  \caption{Frobenius norm of the difference between the ideal core-periphery model and the normalized permuted adjacency matrices (\(\|\Phi_{\text{ideal}} - \Phi_0\|_F\)) as a function of parameters \(\alpha\) and \(\beta\) on Zachary's Karate Club graph for (a) \(\rho = 0.1\), (b) \(\rho = 0.3\), (c) \(\rho = 0.5\), and (d) \(\rho = 0.7\).}
  
    \label{fig:Zachary_Karate_rho}
\end{figure}

\begin{figure}[htbp] 
    \centering
    \begin{subfigure}{0.45\textwidth}
        \centering
        \includegraphics[width=\textwidth]{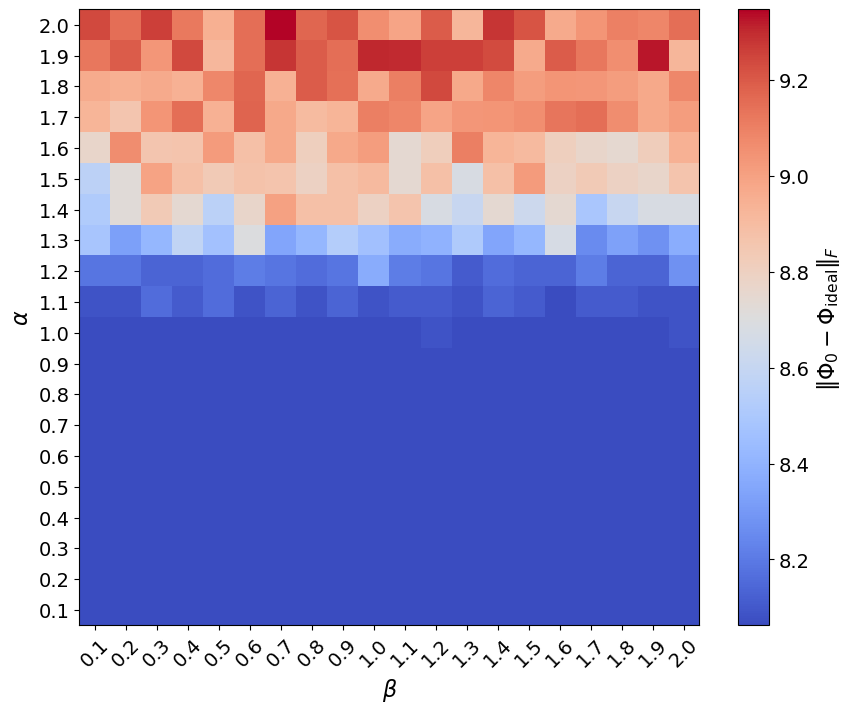}
        \caption{$\rho=0.1$}
        \label{fig:florentine_0_1}
    \end{subfigure}
    \hfill
    \begin{subfigure}{0.45\textwidth}
        \centering
        \includegraphics[width=\textwidth]{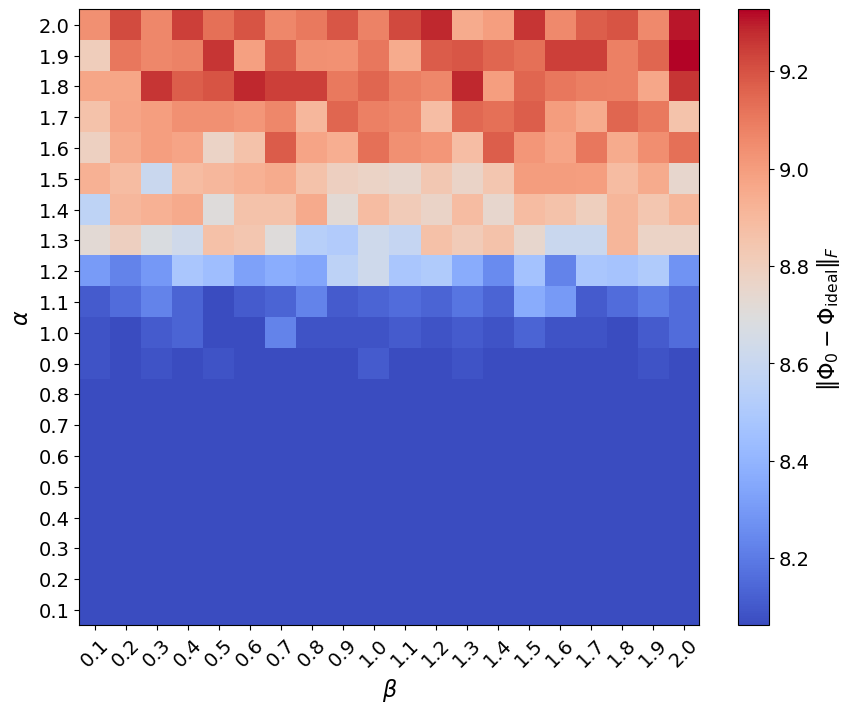}
        \caption{$\rho=0.3$}
        \label{fig:florentine_0_3}
    \end{subfigure}
    
    \vspace{0.5cm} 
    
    \begin{subfigure}{0.45\textwidth}
        \centering
        \includegraphics[width=\textwidth]{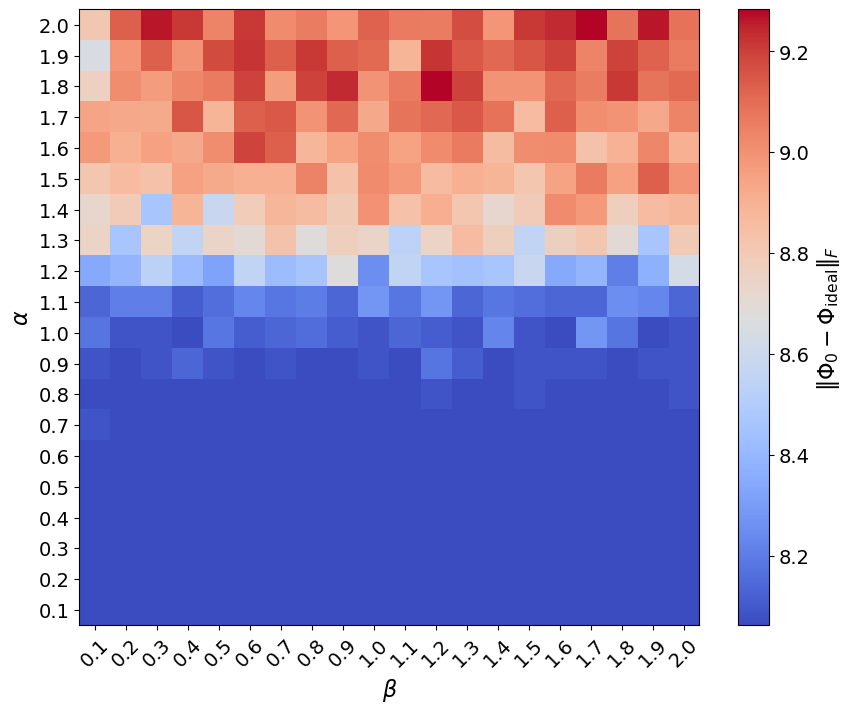}
        \caption{$\rho=0.5$}
        \label{fig:florentine_0_5}
    \end{subfigure}
    \hfill
    \begin{subfigure}{0.45\textwidth}
        \centering
        \includegraphics[width=\textwidth]{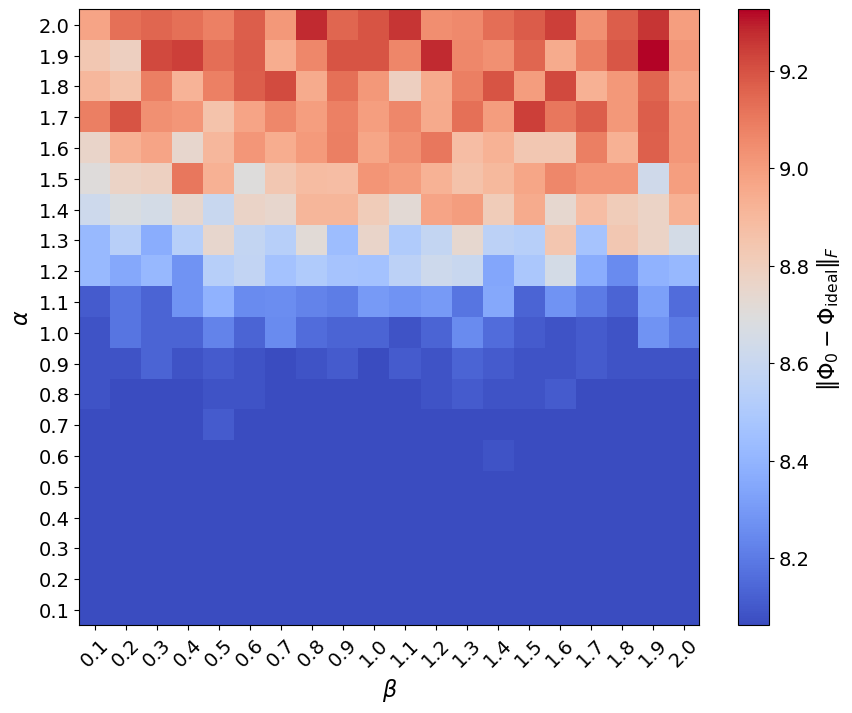}
        \caption{$\rho=0.7$}
        \label{fig:florentine_0_7}
    \end{subfigure}
    
  \caption{Frobenius norm of the difference between the ideal core-periphery model and the normalized permuted adjacency matrices (\(\|\Phi_{\text{ideal}} - \Phi_0\|_F\)) as a function of parameters \(\alpha\) and \(\beta\) on Florentine Families graph for (a) \(\rho = 0.1\), (b) \(\rho = 0.3\), (c) \(\rho = 0.5\), and (d) \(\rho = 0.7\).}
  
    \label{fig:Florentine_Families_rho}
\end{figure}

\begin{figure}[htbp] 
    \centering
    \begin{subfigure}{0.45\textwidth}
        \centering
        \includegraphics[width=\textwidth]{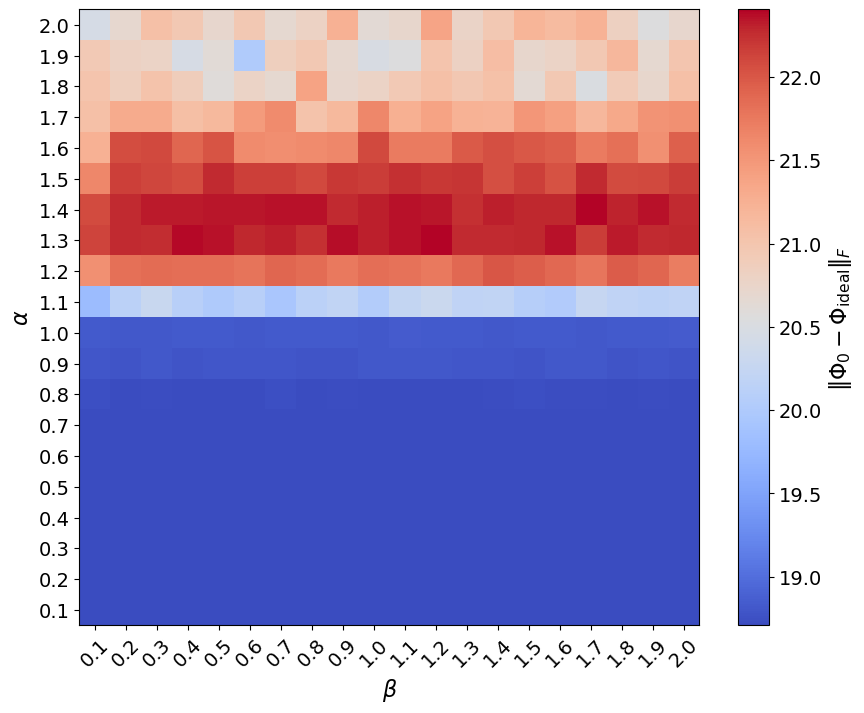}
        \caption{$\rho=0.1$}
        \label{fig:southern_women_0_1}
    \end{subfigure}
    \hfill
    \begin{subfigure}{0.45\textwidth}
        \centering
        \includegraphics[width=\textwidth]{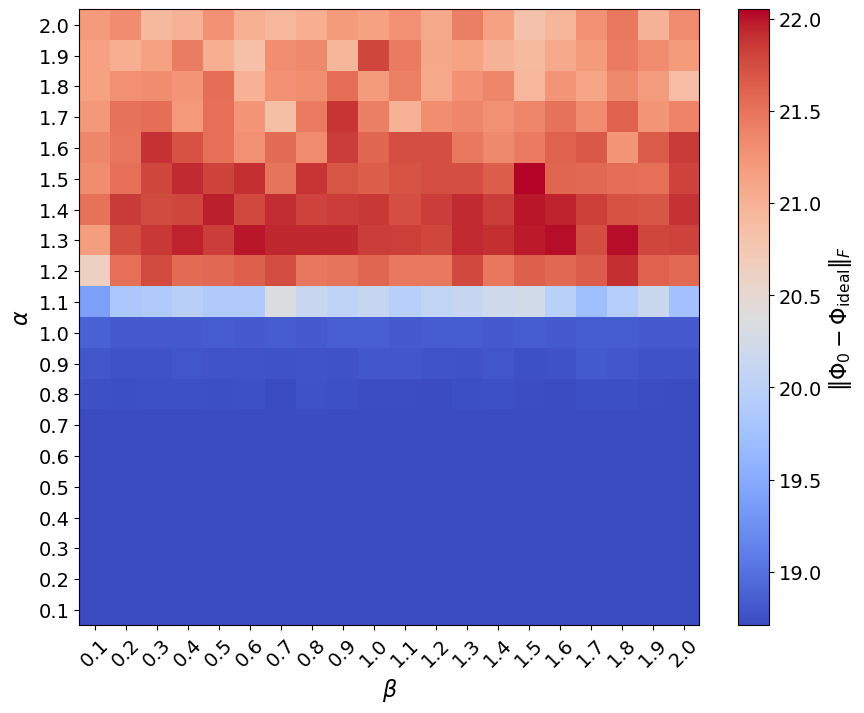}
        \caption{$\rho=0.3$}
        \label{fig:southern_women_0_3}
    \end{subfigure}
    
    \vspace{0.5cm} 
    
    \begin{subfigure}{0.45\textwidth}
        \centering
        \includegraphics[width=\textwidth]{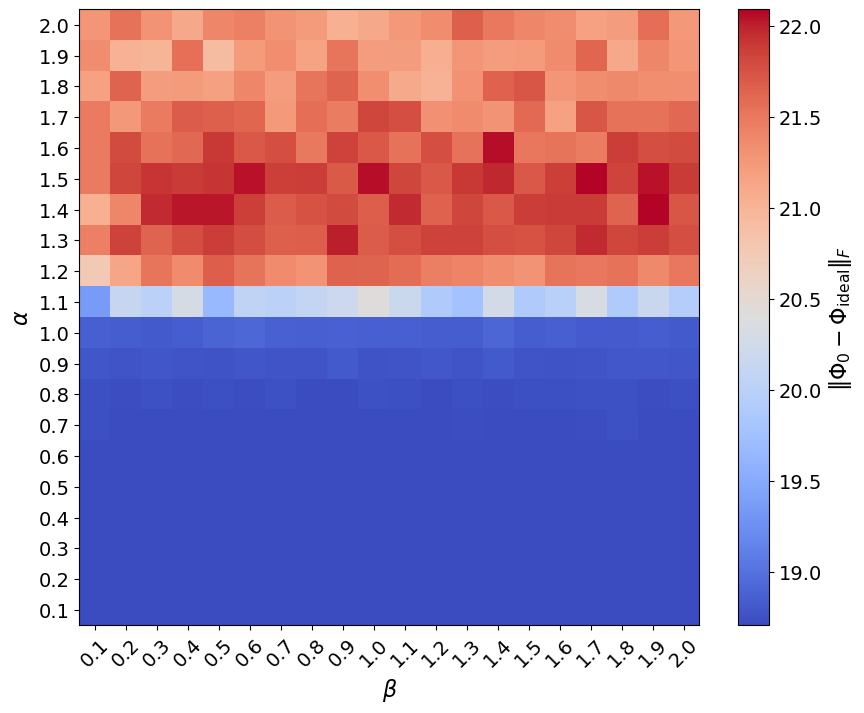}
        \caption{$\rho=0.5$}
        \label{fig:southern_women_0_5}
    \end{subfigure}
    \hfill
    \begin{subfigure}{0.45\textwidth}
        \centering
        \includegraphics[width=\textwidth]{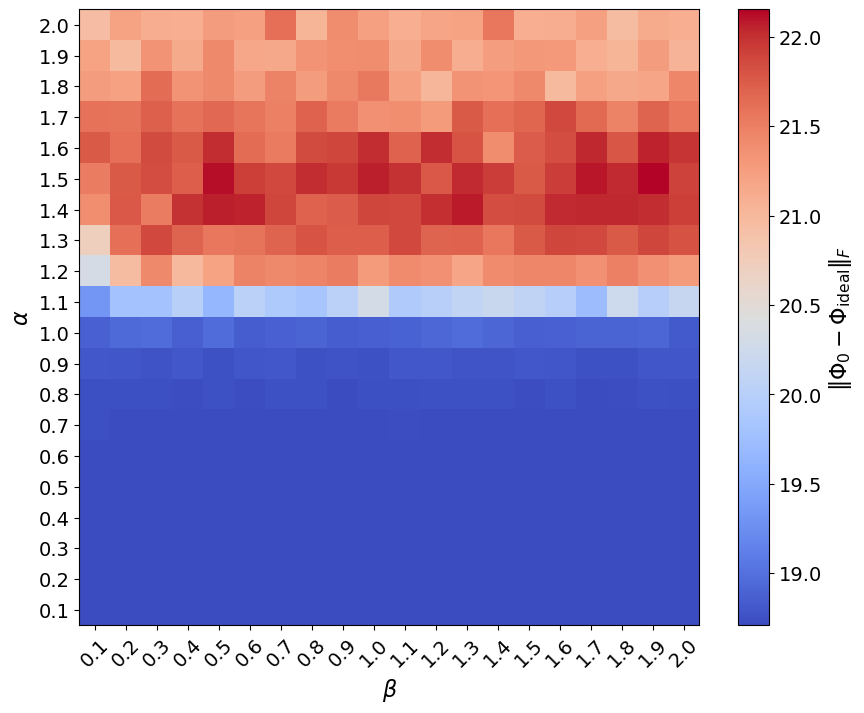}
        \caption{$\rho=0.7$}
        \label{fig:southern_women_0_7}
    \end{subfigure}
    
    \caption{Frobenius norm of the difference between the ideal core-periphery model and the normalized permuted adjacency matrices (\(\|\Phi_{\text{ideal}} - \Phi_0\|_F\)) as a function of parameters \(\alpha\) and \(\beta\) on Davis Southern Women graph for (a) \(\rho = 0.1\), (b) \(\rho = 0.3\), (c) \(\rho = 0.5\), and (d) \(\rho = 0.7\).}
    
    \label{fig:Davis_Southern_Women_rho}
\end{figure}

\begin{figure}[htbp] 
    \centering
    \begin{subfigure}{0.45\textwidth}
        \centering
        \includegraphics[width=\textwidth]{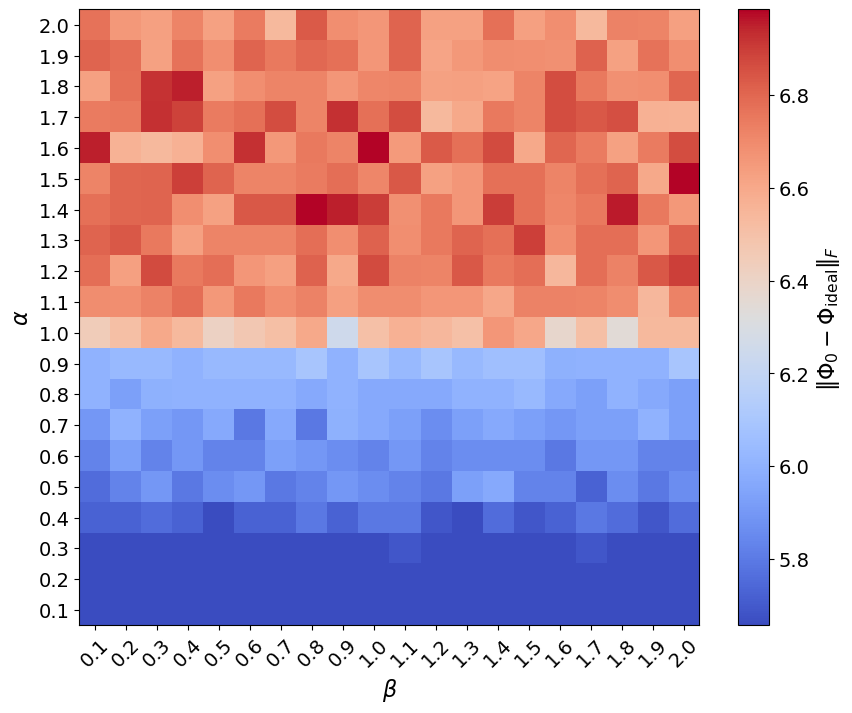}
        \caption{$\rho=0.1$}
        \label{fig:krackhardt_0_1}
    \end{subfigure}
    \hfill
    \begin{subfigure}{0.45\textwidth}
        \centering
        \includegraphics[width=\textwidth]{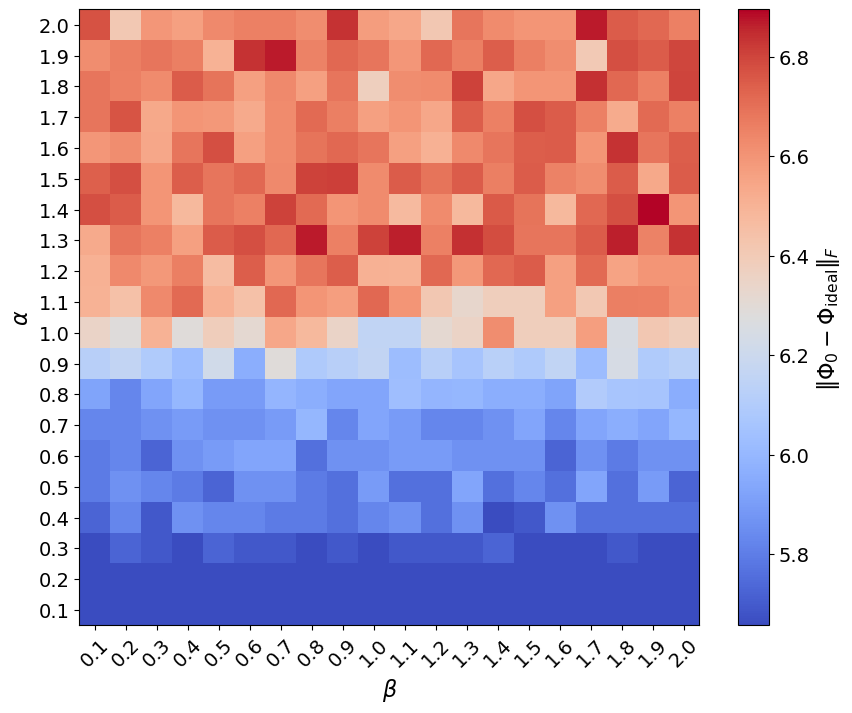}
        \caption{$\rho=0.3$}
        \label{fig:krackhardt_0_3}
    \end{subfigure}
    
    \vspace{0.5cm} 
    
    \begin{subfigure}{0.45\textwidth}
        \centering
        \includegraphics[width=\textwidth]{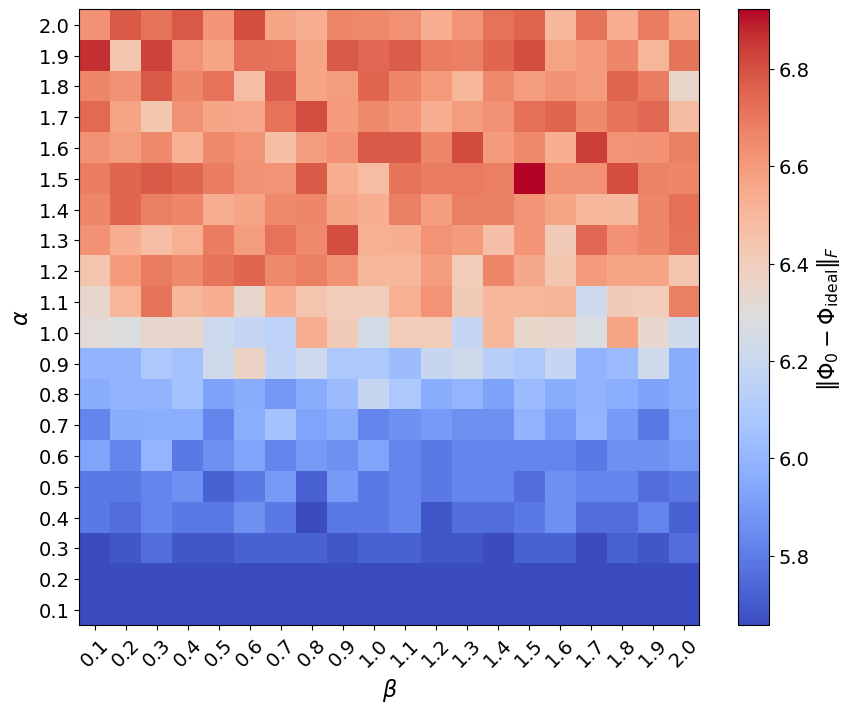}
        \caption{$\rho=0.5$}
        \label{fig:krackhardt_0_5}
    \end{subfigure}
    \hfill
    \begin{subfigure}{0.45\textwidth}
        \centering
        \includegraphics[width=\textwidth]{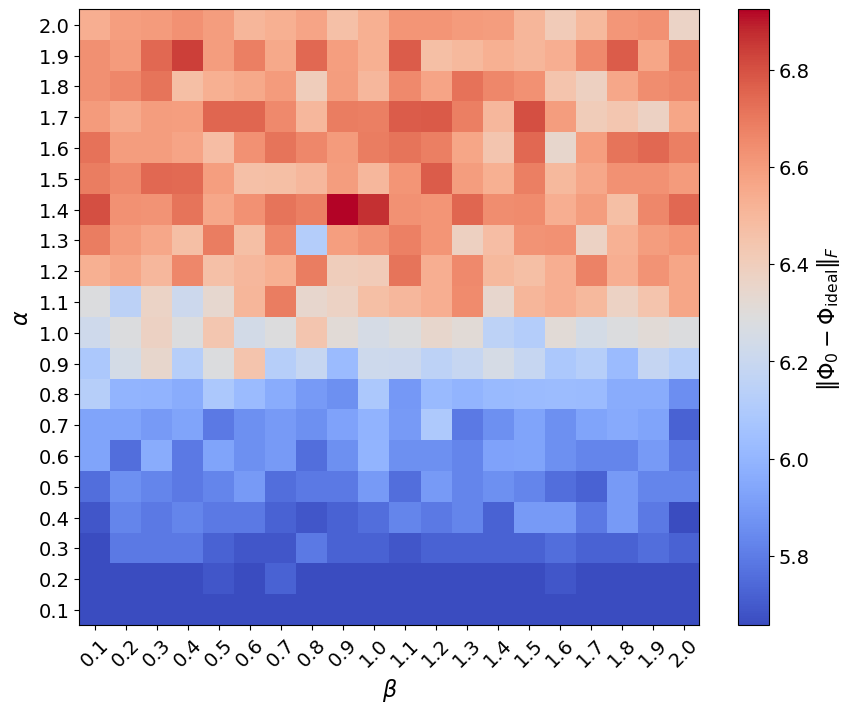}
        \caption{$\rho=0.7$}
        \label{fig:krackhardt_0_7}
    \end{subfigure}
    
    \caption{Frobenius norm of the difference between the ideal core-periphery model and the normalized permuted adjacency matrices (\(\|\Phi_{\text{ideal}} - \Phi_0\|_F\)) as a function of parameters \(\alpha\) and \(\beta\) on the Krackhardt Kite graph for (a) \(\rho = 0.1\), (b) \(\rho = 0.3\), (c) \(\rho = 0.5\), and (d) \(\rho = 0.7\).}
    
    \label{fig:krackhardt_kite_rho}
\end{figure}

The expression for $\Psi_S$, given in equation \ref{eq:psi_S}, incorporates the parameters $\alpha$ and $\beta$, which were calibrated through a sensitivity analysis. To estimate the values of these parameters, along with $\rho$, a detailed analysis was performed. The outcomes of this estimation are presented in \ref{fig:Zachary_Karate_rho}, \ref{fig:Florentine_Families_rho}, \ref{fig:Davis_Southern_Women_rho}, and \ref{fig:krackhardt_kite_rho}. From the figures, it is clear that as the parameters $\alpha$ and $\beta$ increase from their initial values, the Frobenius norm difference between the permuted adjacency matrix and the ideal core-periphery matrix remains relatively stable. However, once $\alpha$ surpasses a threshold of 1, this difference increases for any value of $\beta$. This result is consistent across various values of the evaporation rate, including $\rho = 0.1$, $0.3$, $0.5$, and $0.7$. Therefore, for the rest of the paper, we use the optimum values of \( \alpha = 0.5 \), \( \beta = 1 \) and  \( \beta = 0.5 \).


\FloatBarrier

In the following sub-sections, we describe in detail the comparison between the proposed method and previously developed methods on the various types of graphs: \\


\subsection{Zachary’s Karate Club Network}

 We began our analysis with the well-known Zachary Karate Club network, which provides a depiction of the social connections among 34 members of a university karate club in the United States during the 1970s \cite{zachary1977information}. This network is characterized by 34 vertices and 78 edges, an average degree of 4.59, a density of 0.1390, and an average clustering coefficient of 0.5706. The edges in this network are weighted, reflecting the strength of the connections between the members.

 \FloatBarrier
 
\begin{table}[ht]
\centering
\caption{Top 15 core nodes (Node-ID) of the Zachary’s Karate Club network detected by our proposed method and well-established methods (In brackets we provide their core scores).}
 \label{tab:zachary}
\begin{tabular}{|c|c|c|c|c|}
\hline
Rank & Proposed & Rossa & Rombach & Boyd \\
\hline
1 & 25 (1.00) & 33 (1.00) & 23 (1.00) & 23 (1.00) \\
2 & 2 (0.93) & 0 (0.88) & 33 (0.93) & 31 (0.86) \\
3 & 5 (0.85) & 2 (0.76) & 23 (0.89) & 25 (0.84) \\
4 & 32 (0.78) & 23 (0.67) & 31 (0.86) & 3 (0.83) \\
5 & 27 (0.72) & 1 (0.60) & 25 (0.82) & 13 (0.77) \\
6 & 13 (0.65) & 31 (0.53) & 29 (0.79) & 7 (0.68) \\
7 & 29 (0.58) & 5 (0.46) & 8 (0.25) & 1 (0.67) \\
8 & 33 (0.51) & 3 (0.40) & 27 (0.24) & 0 (0.66) \\
9 & 4 (0.44) & 29 (0.36) & 2 (0.23) & 27 (0.61) \\
10 & 7 (0.37) & 30 (0.30) & 15 (0.22) & 2 (0.57) \\
11 & 12 (0.30) & 24 (0.26) & 30 (0.21) & 29 (0.52) \\
12 & 8 (0.26) & 4 (0.22) & 0 (0.20) & 32 (0.51) \\
13 & 6 (0.21) & 20 (0.18) & 13 (0.19) & 5 (0.50) \\
14 & 31 (0.16) & 14 (0.16) & 26 (0.19) & 6 (0.49) \\
15 & 1 (0.11) & 16 (0.13) & 14 (0.18) & 8 (0.43) \\
\hline
\end{tabular}
\end{table}

Our proposed  ACO-based method was applied to detect core-periphery structures within this network. We compared the performance of our method against several established core-periphery detection techniques, including those developed by Rossa, Rombach, and Boyd et al. To assess the accuracy of the core-periphery structures identified by each method, we used the Frobenius norm to measure the difference between the ideal core-periphery model and the normalized permuted adjacency matrices obtained from each method.


The Frobenius norm difference results for the Zachary Karate Club network were as follows: our proposed method yielded a Frobenius norm difference of 20.212, indicating its performance in aligning with the ideal core-periphery model. In comparison, Rossa's method achieved a norm of 19.927, Rombach's method produced a norm of 20.283, and Boyd resulted in a norm of 20.701. These values suggest that our method performs competitively, with the Frobenius norm difference showing our method's slight advantage over Rombach and Boyd, and a performance close to Rossa.

Additionally, we evaluated the core nodes identified by each method (see Table \ref{tab:zachary}). Our proposed method detected the top core nodes as follows: Node 25 with the highest core score of 1.00, followed by Node 2 with a score of 0.93, and Node 5 with a score of 0.85. Other top core nodes included Node 32 (0.78), Node 27 (0.72), and Node 13 (0.65), among others. In comparison, Rossa's method identified Node 33 as the top core node with a core score of 1.00, while Nodes 0, 2, and 31 were also prominently ranked. Rombach's method listed Node 23 as the top core node, with Nodes 33 and 31 following closely. Boyd identified Node 23 as the highest core node, with Nodes 31 and 25 also ranking highly.


Next, we computed the average cosine similarity results for our proposed method against the methods of Rossa, Rombach, and Boyd et al. by taking 100 realizations. The  cosine similarity between our method and Rossa was 0.666 (SD = 0.060), with the highest value observed between our method and Boyd at 0.679 (SD = 0.033). In comparison, the similarity with Rombach was 0.639 (SD = 0.027), indicating that our method performs competitively with all three comparison algorithms.


\subsection{Florentine Families Network}

Next, we applied our proposed method to the Florentine Families network \cite{breiger1986cumulated}, which captures the marriage relationships among prominent families in Renaissance Florence. In this network, nodes represent families, and edges signify marital ties. The network consists of 16 nodes and 88 edges.

 \FloatBarrier
 
\begin{table}[ht]
\centering
\caption{Top 10 core nodes (Node-ID) of the Florentine Families Network detected by our proposed method and well-established methods (In brackets we provide their core scores).}
 \label{tab:florentine}
\begin{tabular}{|c|c|c|c|c|}
\hline
Rank & Proposed  & Rossa & Rombach & Boyd  \\
\hline
1 & Medici (1.00) & Medici (1.00) & Medici (0.92) & Medici (0.82) \\
2 & Guadagni (0.90) & Strozzi (0.82) & Ridolfi (0.83) & Albizzi (0.65) \\
3 & Strozzi (0.77) & Guadagni (0.67) & Tornabuoni (0.25) & Guadagni (0.57) \\
4 & Peruzzi (0.62) & Peruzzi (0.46) & Strozzi (0.23) & Lamberteschi (0.57) \\
5 & Pazzi (0.50) & Salviati (0.35) & Albizzi (0.21) & Pazzi (0.57) \\
6 & Ginori (0.39) & Barbadori (0.29) & Barbadori (0.19) & Peruzzi (0.56) \\
7 & Tornabuoni (0.27) & Ridolfi (0.21) & Acciaiuoli (0.17) & Strozzi (0.49) \\
8 & Barbadori (0.15) & Ginori (0.12) & Salviati (0.15) & Tornabuoni (0.49) \\
9 & Acciaiuoli (0.00) & Acciaiuoli (0.00) & Castellani (0.12) & Ridolfi (0.47) \\
10 & Castellani (0.00) & Castellani (0.00) & Guadagni (0.10) & Bischeri (0.45) \\
\hline
\end{tabular}
\end{table}


 The The differences in Frobenius norm differences between the ideal core-periphery structure and the obtained structure were as follows: our proposed method achieved a norm of 8.30, Rossa's method produced a norm of 8.307, Rombach's method yielded a norm of 9.220, and Boyd resulted in a norm of 10.247. These values indicate that our method performs competitively and slightly better than the established methods, with the lowest Frobenius norm difference reflecting a closer approximation to the ideal core-periphery structure.

In terms of identifying the top core nodes within the Florentine Families network (see Table \ref{tab:florentine}), our proposed method detected the following top 10 nodes: Medici with the highest core score of 1.00, Guadagni with a score of 0.90, and Strozzi with a score of 0.77. Other notable nodes included Peruzzi (0.62), Pazzi (0.50), and Ginori (0.39). Rossa's method identified Medici as the top core node with a core score of 1.00, followed by Strozzi (0.82) and Guadagni (0.67). Rombach's method ranked Medici as the top core node (0.92), with Ridolfi (0.83) and Tornabuoni (0.25) also featured prominently. Boyd found Medici as the leading core node (0.82), with Albizzi (0.65) and Guadagni (0.57) among the top nodes.


 For the Florentine Families network, the cosine similarity results show that our proposed method achieved a similarity of 0.705 (SD = 0.163) with Rossa, indicating a high degree of similarity with some variability. The similarity with Rombach was 0.560 (SD = 0.026), and with Boyd, it was 0.644 (SD = 0.024). 
 

\subsection{Davis Southern Women Network}
The Davis Southern Women network records social events attended by women in the American South during the 1930s. Nodes correspond to women, and edges indicate shared event attendance \cite{davis1941deep}. The network consists of 32 nodes and 89 edges. The The differences in Frobenius norm difference between the ideal core-periphery structure and the obtained structures were as follows: obtained were as follows: our proposed method achieved a norm of 18.815, Rossa's method resulted in a norm of 19.339, Rombach's method yielded a norm of 20.640, and Boyd produced a norm of 21.024. These values indicate that our method performs competitively, with the lowest Frobenius norm difference reflecting a closer approximation to the ideal core-periphery structure.

 
 \FloatBarrier

\begin{table}[ht]
\centering
\caption{Top 15 core nodes of the Davis Southern women social network detected by our proposed method and well-established methods (in brackets we provide their core scores).}
\label{tab:davis}
\resizebox{\textwidth}{!}{ 
\begin{tabular}{|c|c|c|c|c|}
\hline
Rank & Proposed & Rossa & Rombach & Boyd \\
\hline
1 & E8 (1.00) & E8 (1.00) & Theresa Anderson (0.96) & E8 (0.84) \\
2 & E5 (0.96) & E9 (0.91) & E8 (0.93) & E9 (0.69) \\
3 & E9 (0.92) & E7 (0.83) & E6 (0.89) & E6 (0.68) \\
4 & E7 (0.87) & E6 (0.75) & Evelyn Jefferson (0.86) & Pearl Oglethorpe (0.68) \\
5 & E6 (0.82) & E5 (0.67) & E5 (0.82) & Dorothy Murchison (0.64) \\
6 & E12 (0.77) & E3 (0.59) & Brenda Rogers (0.79) & E13 (0.63) \\
7 & E11 (0.71) & Nora Fayette (0.52) & E3 (0.25) & E5 (0.61) \\
8 & E10 (0.64) & E4 (0.46) & Laura Mandeville (0.24) & Nora Fayette (0.60) \\
9 & E3 (0.57) & Katherina Rogers (0.41) & E9 (0.23) & E12 (0.57) \\
10 & E4 (0.49) & Sylvia Avondale (0.35) & Pearl Oglethorpe (0.22) & Sylvia Avondale (0.54) \\
11 & E2 (0.41) & E1 (0.29) & Frances Anderson (0.21) & Eleanor Nye (0.54) \\
12 & E14 (0.32) & E2 (0.24) & Eleanor Nye (0.20) & Charlotte McDowd (0.51) \\
13 & E13 (0.23) & Helen Lloyd (0.18) & E7 (0.19) & E7 (0.51) \\
14 & E1 (0.12) & E12 (0.12) & Nora Fayette (0.18) & Olivia Carleton (0.51) \\
15 & Evelyn Jefferson (0.00) & Olivia Carleton (0.08) & Ruth DeSand (0.17) & E14 (0.51) \\
\hline
\end{tabular}
}
\end{table}

In terms of identifying the top core nodes within the Davis Southern Women network(see Table \ref{tab:davis}), our proposed method detected the following top 15 nodes: E8 with a core score of 1.00, E5 with a score of 0.96, and E9 with a score of 0.92. Other notable nodes included E7 (0.87), E6 (0.82), and E12 (0.77). Rossa's method identified E8 as the top core node (1.00), followed by E5 (0.96) and E9 (0.91). Rombach's method found Theresa Anderson as the top core node (0.96), with E8 (0.93) and E6 (0.89) also ranking highly. Boyd identified E8 as the leading core node (0.84), with E9 (0.69) and E6 (0.68) among the top nodes.

The cosine similarity between our proposed method and Rossa was 0.813 (SD=0.252), reflecting a high degree of similarity with some variability. The similarity between our method and Rombach was 0.550 (SD=0.019), and between our method and Boyd was 0.628 (SD=0.029).


\subsection{Les Miserables Network}

This network examines the co-occurrence of characters in Victor Hugo's \textit{Les Miserables} \cite{knuth1993stanford}. Nodes represent characters, and edges reflect their interactions within the novel. The Les Miserables network is characterized by 77 vertices and 254 edges, an average degree of 6.60, a density of 0.0864, and an average clustering coefficient of 0.5733. The network is unweighted.

 \FloatBarrier
 
\begin{table}[ht]
\centering
\caption{Top 20 core nodes of the Les Miserables network detected by our proposed method and well-established methods (in brackets we provide their core scores).}
\label{tab:les_Miserables}
\resizebox{\textwidth}{!}{ 
\begin{tabular}{|c|c|c|c|c|}
\hline
Rank & Proposed & Rossa & Rombach & Boyd \\
\hline
1 & Valjean (1.00) & Valjean (1.00) & Enjolras (0.98) & Thenardier (1.00) \\
2 & Marius (0.92) & Enjolras (0.89) & Valjean (0.97) & Cosette (0.85) \\
3 & Enjolras (0.85) & Marius (0.83) & Marius (0.95) & Gavroche (0.81) \\
4 & Thenardier (0.78) & Bossuet (0.77) & Courfeyrac (0.94) & Courfeyrac (0.77) \\
5 & MmeMagloire (0.73) & Thenardier (0.73) & Combeferre (0.92) & Combeferre (0.75) \\
6 & MlleGillenormand (0.68) & Gavroche (0.69) & Cosette (0.91) & Javert (0.74) \\
7 & Joly (0.65) & Blacheville (0.66) & Thenardier (0.89) & MmeThenardier (0.74) \\
8 & Courfeyrac (0.61) & Myriel (0.62) & Bossuet (0.88) & Bahorel (0.69) \\
9 & Dahlia (0.57) & Dahlia (0.59) & Javert (0.86) & Fantine (0.66) \\
10 & Blacheville (0.54) & Combeferre (0.57) & Feuilly (0.84) & Marius (0.59) \\
11 & Babet (0.50) & Gueulemer (0.54) & Bahorel (0.83) & Joly (0.59) \\
12 & Gavroche (0.47) & Listolier (0.51) & Joly (0.81) & Babet (0.56) \\
13 & Fantine (0.44) & Joly (0.49) & Fantine (0.80) & Bossuet (0.55) \\
14 & Champmathieu (0.41) & MlleGillenormand (0.46) & Gavroche (0.78) & Enjolras (0.54) \\
15 & Tholomyes (0.39) & Zephine (0.44) & MmeThenardier (0.77) & Gillenormand (0.51) \\
16 & Feuilly (0.37) & Judge (0.42) & Babet (0.25) & Gueulemer (0.47) \\
17 & Claquesous (0.35) & Javert (0.40) & Eponine (0.25) & Valjean (0.44) \\
18 & MotherInnocent (0.33) & Bahorel (0.38) & Gueulemer (0.24) & Feuilly (0.44) \\
19 & Child2 (0.31) & Eponine (0.36) & Grantaire (0.24) & Claquesous (0.43) \\
20 & MotherPlutarch (0.29) & Fameuil (0.34) & Fauchelevent (0.23) & Favourite (0.41) \\
\hline
\end{tabular}
}
\end{table}


The Frobenius norm difference results for the Les Miserables network were as follows: Our proposed method achieved a norm of 49.910, indicating a close approximation to the ideal core-periphery model. In comparison, Rossa's method produced a norm of 49.851, Rombach's method resulted in a norm of 49.935, and Boyd yielded a norm of 50.066. These values suggest that our method performs competitively, with the Frobenius norm difference showing that our method is slightly better aligned with the ideal core-periphery model than Boyd and comparable to Rossa and Rombach.

Table \ref{tab:les_Miserables} presents a comparison of the top 20 core nodes detected by our proposed method, Rossa, Rombach, and Boyd. Our method, like Rossa, ranked Valjean as the top core node (1.00), but with notable differences in subsequent rankings. For example, Marius was ranked second (0.92) by our method, while Rossa ranked him third (0.83), and Rombach placed him in the second position (0.95). Boyd, on the other hand, ranked Thenardier as the top node (1.00), with Cosette (0.85) and Gavroche (0.81) appearing among the top nodes, differing from our method's focus on Marius and Enjolras. Additionally, our method identified MmeMagloire (0.73) and Joly (0.65) among the top nodes, whereas Rossa placed them lower, and Rombach identified Courfeyrac (0.94) and Cosette (0.91) as top nodes, showing consistent performance but different preferences for specific nodes.

 For the Les Miserables network, the cosine similarity between our proposed method and Rossa was 0.784 (SD = 0.070), indicating a high degree of similarity with some variability. The similarity between our method and Rombach was 0.714 (SD = 0.005), and between our method and Boyd was 0.689 (SD = 0.022).


\subsection{Word Adjacencies}
This adjacency network consists of common adjectives and nouns found in the novel "David Copperfield" by Charles Dickens \cite{newman2006finding}. The network is composed of 112 vertices and 425 edges, with an average degree of 7.59, a density of 0.0684, and an average clustering coefficient of 0.1728. The network is unweighted.

 \FloatBarrier
 
\begin{table}[ht]
\centering
\caption{Top 30 core nodes (Node-ID) of the Word adjacencies network detected by our proposed method and well-established methods (In brackets we provide their core scores).}
\label{tab:word_adjacencies}
\begin{tabular}{|c|c|c|c|c|}
\hline
Rank & Proposed & Rossa & Rombach & Boyd \\
\hline
1 & 50 (1) & 2 (1) & 17 (0.98913) & 51 (0.64837) \\
2 & 25 (0.98799) & 43 (0.95961) & 51 (0.97826) & 21 (0.63134) \\
3 & 3 (0.97570) & 51 (0.92269) & 2 (0.96739) & 102 (0.63019) \\
4 & 1 (0.96311) & 104 (0.88568) & 43 (0.95652) & 71 (0.62490) \\
5 & 75 (0.95024) & 9 (0.85676) & 104 (0.94565) & 98 (0.62484) \\
6 & 87 (0.93705) & 28 (0.83264) & 50 (0.93478) & 62 (0.61872) \\
7 & 11 (0.92355) & 79 (0.81127) & 25 (0.92391) & 2 (0.61834) \\
8 & 89 (0.90974) & 76 (0.78910) & 12 (0.91304) & 86 (0.61441) \\
9 & 71 (0.89569) & 70 (0.76788) & 9 (0.90217) & 91 (0.61122) \\
10 & 29 (0.88130) & 50 (0.74889) & 18 (0.89130) & 23 (0.61012) \\
11 & 31 (0.86672) & 37 (0.72948) & 1 (0.88043) & 43 (0.60881) \\
12 & 5 (0.85191) & 27 (0.70988) & 21 (0.86957) & 41 (0.58721) \\
13 & 38 (0.83678) & 25 (0.69194) & 3 (0.85870) & 101 (0.58698) \\
14 & 42 (0.82145) & 1 (0.67528) & 102 (0.84783) & 33 (0.57594) \\
15 & 81 (0.80611) & 24 (0.65785) & 26 (0.83696) & 82 (0.55980) \\
16 & 106 (0.79052) & 7 (0.64068) & 24 (0.82609) & 80 (0.54493) \\
17 & 72 (0.77454) & 14 (0.62522) & 70 (0.81522) & 8 (0.54454) \\
18 & 33 (0.75839) & 65 (0.61023) & 54 (0.80435) & 72 (0.53940) \\
19 & 47 (0.74236) & 54 (0.59537) & 31 (0.79348) & 111 (0.53888) \\
20 & 26 (0.72605) & 15 (0.58029) & 59 (0.78261) & 89 (0.51436) \\
21 & 20 (0.70952) & 71 (0.56716) & 41 (0.77174) & 77 (0.51344) \\
22 & 34 (0.69265) & 12 (0.55472) & 27 (0.76087) & 17 (0.50375) \\
23 & 22 (0.67559) & 5 (0.54231) & 68 (0.25) & 73 (0.49477) \\
24 & 92 (0.65812) & 72 (0.53021) & 7 (0.24719) & 14 (0.49134) \\
25 & 102 (0.64051) & 52 (0.51779) & 14 (0.24438) & 74 (0.48159) \\
26 & 55 (0.62294) & 11 (0.50501) & 19 (0.24157) & 103 (0.47365) \\
27 & 27 (0.60534) & 89 (0.49187) & 15 (0.23876) & 6 (0.46359) \\
28 & 65 (0.58727) & 31 (0.47835) & 66 (0.23596) & 70 (0.45560) \\
29 & 15 (0.57031) & 23 (0.46512) & 32 (0.23315) & 78 (0.45226) \\
30 & 21 (0.55307) & 3 (0.45161) & 36 (0.23034) & 53 (0.43757) \\
\hline
\end{tabular}
\end{table}

 \FloatBarrier

The Frobenius norm difference results for the Word Adjacencies network were as follows: The results in Table \ref{tab:frobenius_norm_difference} highlight that our proposed method achieved a norm of 73.553, indicating a close approximation to the ideal core-periphery model. In comparison, Rossa's method produced a norm of 70.640, Rombach's method resulted in a norm of 71.091, and Boyd yielded a norm of 73.959. These values suggest that our method performs competitively, with the Frobenius norm difference showing a similar alignment with the ideal core-periphery model as Rossa and Rombach, and a slightly better performance than Boyd.

Table \ref{tab:word_adjacencies} compares the top 30 core nodes detected by our proposed method, Rossa, Rombach, and Boyd for the Word Adjacencies network. Node 50 (1.00) ranks first in our method, while Rossa ranks Node 2 (1.00), Rombach places Node 17 (0.98913), and Boyd ranks Node 51 (0.64837). Our method also places Node 25 (0.98799) second, while Rossa and Rombach rank Nodes 43 (0.95961) and 51 (0.97826), respectively, in the second position. Boyd places Node 21 (0.63134) second. At lower rankings, variations are observed, such as Node 22, which is ranked 23rd by our method but much lower by Rombach and Boyd. These discrepancies suggest varying node prioritization across methods.

For the Word Adjacencies network, the cosine similarity between our proposed method and Rossa was 0.555 (SD = 0.071), indicating moderate similarity. The similarity between our method and Rombach was 0.594 (SD = 0.015), and between our method and Boyd was 0.594 (SD = 0.032).


\subsection{Dolphin Social Network}
This network captures the frequent associations among 62 dolphins in a community residing in Doubtful Sound, New Zealand \cite{lusseau2003bottlenose}. The network is composed of 62 vertices and 159 edges, with an average degree of 5.13, a density of 0.0841, and an average clustering coefficient of 0.2590. The network is unweighted.

 \FloatBarrier
 
\begin{table}[ht]
\centering
\caption{Top 30 core nodes of the Dolphin social network network detected by our proposed method and well-established methods (In brackets we provide their core scores).}
\label{tab:dolphin_social}
\begin{tabular}{|c|c|c|c|c|}
\hline
Rank & Proposed & Rossa & Rombach & Boyd \\
\hline
1 & 51 (1.00) & 45 (1.00) & 45 (0.98) & 51 (0.66) \\
2 & 14 (0.98) & 33 (0.96) & 33 (0.96) & 1 (0.66) \\
3 & 36 (0.96) & 20 (0.93) & 18 (0.94) & 53 (0.64) \\
4 & 15 (0.93) & 57 (0.90) & 37 (0.92) & 60 (0.63) \\
5 & 9 (0.91) & 1 (0.86) & 21 (0.90) & 36 (0.62) \\
6 & 54 (0.88) & 37 (0.83) & 24 (0.88) & 23 (0.62) \\
7 & 33 (0.86) & 29 (0.79) & 51 (0.87) & 2 (0.60) \\
8 & 30 (0.83) & 0 (0.76) & 29 (0.85) & 56 (0.60) \\
9 & 10 (0.81) & 17 (0.73) & 14 (0.83) & 37 (0.60) \\
10 & 27 (0.78) & 38 (0.69) & 50 (0.81) & 25 (0.60) \\
11 & 20 (0.75) & 30 (0.67) & 40 (0.79) & 42 (0.58) \\
12 & 45 (0.72) & 54 (0.64) & 15 (0.77) & 44 (0.51) \\
13 & 57 (0.69) & 9 (0.61) & 16 (0.25) & 35 (0.50) \\
14 & 49 (0.66) & 2 (0.58) & 43 (0.24) & 58 (0.48) \\
15 & 56 (0.63) & 40 (0.56) & 34 (0.24) & 32 (0.46) \\
16 & 38 (0.60) & 59 (0.53) & 20 (0.23) & 46 (0.45) \\
17 & 53 (0.57) & 24 (0.50) & 38 (0.23) & 39 (0.45) \\
18 & 29 (0.54) & 50 (0.47) & 8 (0.22) & 50 (0.44) \\
19 & 37 (0.51) & 13 (0.45) & 0 (0.22) & 29 (0.40) \\
20 & 17 (0.47) & 27 (0.42) & 7 (0.21) & 20 (0.38) \\
21 & 13 (0.44) & 34 (0.40) & 59 (0.21) & 3 (0.36) \\
22 & 8 (0.41) & 47 (0.37) & 10 (0.20) & 45 (0.36) \\
23 & 42 (0.37) & 18 (0.35) & 61 (0.20) & 43 (0.36) \\
24 & 26 (0.34) & 43 (0.32) & 55 (0.19) & 14 (0.33) \\
25 & 44 (0.31) & 56 (0.30) & 23 (0.19) & 54 (0.33) \\
26 & 40 (0.28) & 55 (0.27) & 54 (0.18) & 33 (0.32) \\
27 & 60 (0.25) & 23 (0.25) & 36 (0.18) & 17 (0.30) \\
28 & 19 (0.21) & 3 (0.22) & 12 (0.17) & 38 (0.30) \\
29 & 47 (0.19) & 11 (0.20) & 57 (0.17) & 21 (0.29) \\
30 & 59 (0.17) & 49 (0.18) & 52 (0.16) & 18 (0.29) \\
\hline
\end{tabular}
\end{table}

 \FloatBarrier
 
The Frobenius norm difference results for the Dolphin Social Network were as follows: Our proposed method achieved a norm of 39.154, indicating a close approximation to the ideal core-periphery model. In comparison, Rossa's method produced a norm of 38.536, Rombach's method resulted in a norm of 40.361, and Boyd yielded a norm of 41.725. These values suggest that our method performs competitively, with the Frobenius norm difference showing a similar alignment with the ideal core-periphery model as Rossa and Rombach, and a slightly better performance than Boyd.

In terms of identifying the top core nodes (see Table \ref{tab:dolphin_social}), our proposed method detected the following top core nodes: Node 51 with a core score of 1.00, followed by Node 14 with a score of 0.98, and Node 36 with a score of 0.96. Other notable nodes included Node 15 (0.93), Node 9 (0.91), and Node 54 (0.88). In comparison, Rossa's method identified Node 45 as the top core node with a core score of 1.00, while Nodes 33, 20, and 57 were also prominently ranked. Rombach's method listed Node 45 as the top core node (0.98), with Nodes 33 and 18 following closely. Boyd found Node 51 as the leading core node (0.66), with Nodes 1 and 53 also among the top-ranked nodes.


For the Dolphin Social Network, the cosine similarity between our proposed method and Rossa was 0.709 (SD = 0.056), indicating moderate to high similarity. The similarity between our method and Rombach was 0.571 (SD = 0.020), and between our method and Boyd was 0.616 (SD = 0.043).


 \subsection{Books About US Politics}
This network consists of books on US politics that were published around the time of the 2004 presidential election and sold by Amazon.com. The edges between books indicate frequent copurchasing by the same customers. The network is composed of 105 vertices and 441 edges, with an average degree of 8.40, a density of 0.0808, and an average clustering coefficient of 0.4875. The network is unweighted.

 \FloatBarrier

\begin{table}[ht]
\centering
\caption{Top 30 core nodes (Node-ID) of the books about US politics network detected by our proposed method and well-established methods (In brackets we provide their core scores).}
\label{tab:books_about_US}
\begin{tabular}{|c|c|c|c|c|}
\hline
Rank & Proposed & Rossa & Rombach & Boyd \\
\hline

1 & 9 (0.9883) & 12 (0.6305) & 72 (0.1815) & 13 (0.3374) \\
2 & 30 (0.9763) & 84 (0.8560) & 30 (0.9643) & 38 (0.3376) \\
3 & 6 (0.9641) & 72 (0.6516) & 66 (0.0923) & 58 (0.2090) \\
4 & 13 (0.9515) & 3 (0.0000) & 84 (0.1339) & 102 (0.6684) \\
5 & 35 (0.9387) & 30 (0.4971) & 74 (0.1220) & 12 (0.5690) \\
6 & 64 (0.9258) & 47 (0.5318) & 75 (0.2113) & 42 (0.5416) \\
7 & 100 (0.9126) & 73 (0.5694) & 86 (0.8929) & 47 (0.2590) \\
8 & 72 (0.9122) & 8 (0.9762) & 73 (0.5828) & 73 (0.9762) \\
9 & 23 (0.8992) & 66 (0.4009) & 100 (0.1280) & 54 (0.5669) \\
10 & 71 (0.8854) & 40 (0.7237) & 76 (0.8690) & 8 (0.3554) \\
11 & 94 (0.8715) & 71 (0.4306) & 71 (0.2321) & 72 (0.3679) \\
12 & 26 (0.8574) & 11 (0.2362) & 82 (0.1369) & 104 (0.2007) \\
13 & 61 (0.8430) & 58 (0.2174) & 79 (0.1399) & 82 (0.1005) \\
14 & 58 (0.8290) & 6 (0.6761) & 99 (0.1845) & 5 (0.6505) \\
15 & 104 (0.8149) & 9 (0.3239) & 83 (0.0030) & 77 (0.5797) \\
16 & 20 (0.8006) & 75 (0.0000) & 31 (0.1429) & 35 (0.4706) \\
17 & 75 (0.7860) & 10 (0.6105) & 93 (0.9167) & 23 (0.2823) \\
18 & 56 (0.7710) & 100 (0.1092) & 91 (0.0536) & 28 (0.0579) \\
19 & 97 (0.7557) & 74 (0.1744) & 89 (0.2232) & 95 (0.4329) \\
20 & 4 (0.7404) & 64 (0.4790) & 70 (0.1726) & 103 (0.1900) \\
21 & 74 (0.7248) & 86 (0.5491) & 98 (0.9286) & 93 (0.3562) \\
22 & 63 (0.7095) & 35 (0.3121) & 77 (0.1250) & 90 (0.0371) \\
23 & 39 (0.6940) & 4 (0.2468) & 87 (0.0833) & 64 (0.2568) \\
24 & 41 (0.6780) & 57 (0.2569) & 78 (0.1071) & 96 (0.2802) \\
25 & 3 (0.6617) & 59 (0.8844) & 90 (0.1964) & 21 (0.5071) \\
26 & 55 (0.6452) & 94 (0.2258) & 62 (0.0060) & 66 (0.4841) \\
27 & 65 (0.6284) & 65 (0.4160) & 94 (0.1012) & 40 (0.2772) \\
28 & 76 (0.6113) & 23 (0.0193) & 49 (0.8810) & 3 (0.3191) \\
29 & 102 (0.5943) & 34 (0.3007) & 60 (0.1190) & 10 (0.6348) \\
30 & 90 (0.5767) & 53 (0.1834) & 97 (0.2381) & 55 (0.5425) \\
\hline
\end{tabular}
\end{table}

 \FloatBarrier

The Frobenius norm difference results for the Books About US Politics network were as follows: our proposed method, as shown in Table \ref{tab:frobenius_norm_difference}, achieved a norm of 68.891, indicating a close approximation to the ideal core-periphery model. In comparison, Rossa's method produced a norm of 65.833, Rombach's method resulted in a norm of 70.271, and Boyd yielded a norm of 70.441. These values suggest that our method performs competitively, with the Frobenius norm difference showing a similar alignment with the ideal core-periphery model as Rossa, and a slightly better performance than Rombach and Boyd.

In terms of identifying the top core nodes, our proposed method detected the following top core nodes (see Table \ref{tab:books_about_US}): Node 72 with a core score of 0.9122, followed by Node 9 with a score of 0.9883, and Node 30 with a score of 0.9763. Other notable nodes included Node 6 (0.9641), Node 13 (0.9515), and Node 35 (0.9387). In comparison, Rossa's method identified Node 8 as the top core node with a core score of 0.9762, while Nodes 12, 84, and 3 were also prominently ranked. Rombach's method listed Node 73 as the top core node (0.5828), with Nodes 72 and 30 following closely. Boyd found Node 73 as the leading core node (0.9762), with Nodes 13 and 102 also among the top-ranked nodes.


 For the Books About US Politics network, the cosine similarity between our proposed method and Rossa was 0.746 with a standard deviation of 0.030, indicating high similarity. The similarity between our method and Rombach was 0.562 (SD = 0.004), and between our method and Boyd was 0.641 (SD = 0.034). \\

\begin{figure}[htbp]
    \centering

    \begin{subfigure}{0.85\textwidth}
        \centering
        \includegraphics[width=\textwidth]{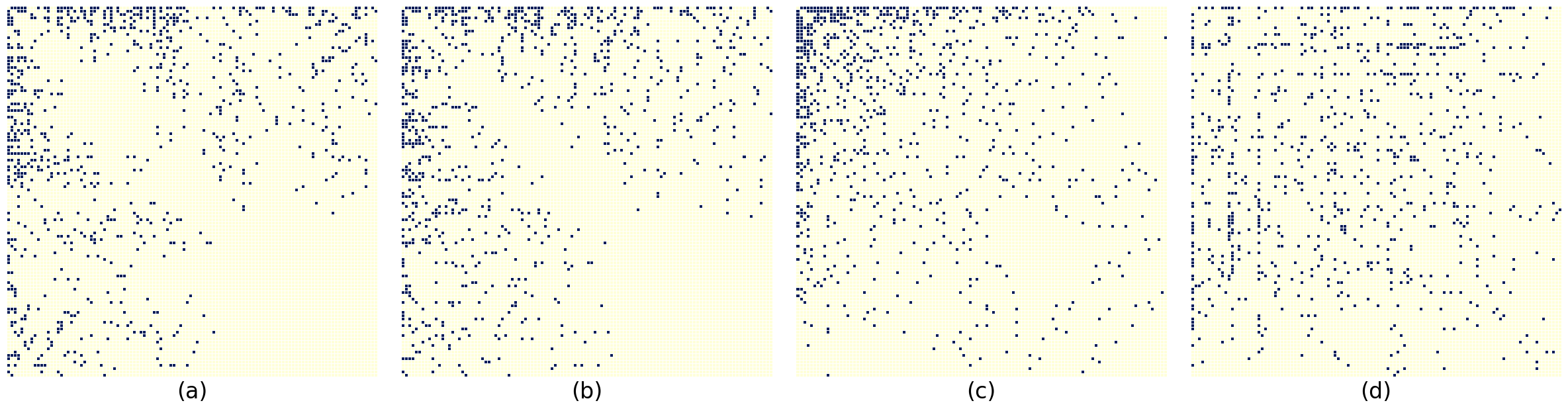}
        \caption*{(A) Word Adjacencies}
        \label{fig:permuted_adjacency_adjnoun}
    \end{subfigure}
    
    \vspace{0.5cm} 
    
    \begin{subfigure}{0.85\textwidth}
        \centering
        \includegraphics[width=\textwidth]{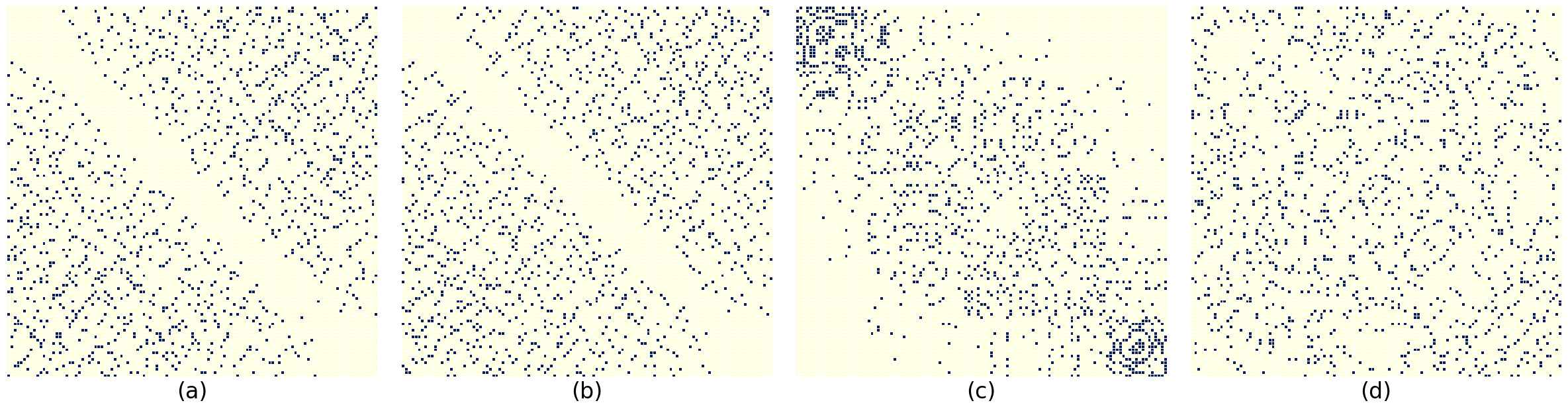}
        \caption*{(B) American College Football}
        \label{fig:permuted_adjacency_football}
    \end{subfigure}
    
    \vspace{0.5cm} 

    \begin{subfigure}{0.85\textwidth}
        \centering
        \includegraphics[width=\textwidth]{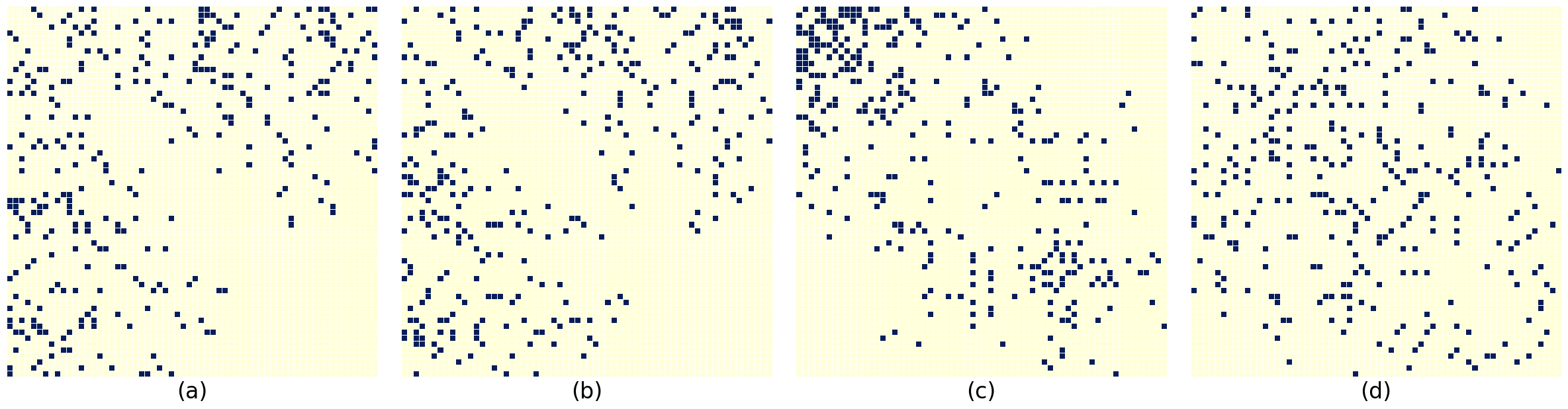}
        \caption*{(C) Dolphins}
        \label{fig:permuted_adjacency_dolphins}
    \end{subfigure}
    
    \vspace{0.5cm} 

    \begin{subfigure}{0.85\textwidth}
        \centering
        \includegraphics[width=\textwidth]{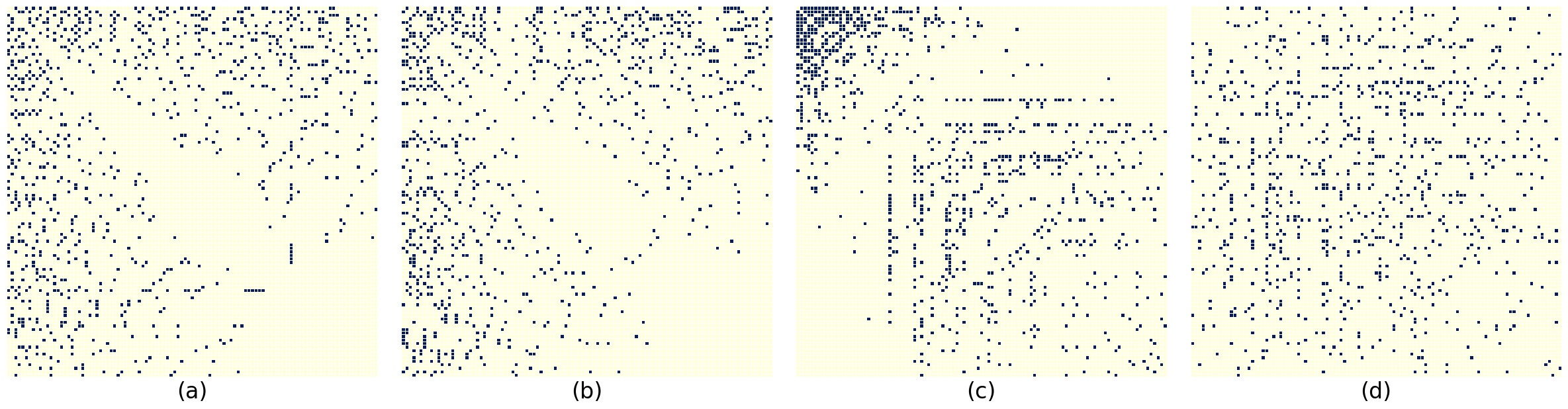}
        \caption*{(D) Books About US Politics}
        \label{fig:permuted_adjacency_polbooks}
    \end{subfigure}

  \caption{Groundtruth adjacency matrices for the networks: (A) Word Adjacencies, (B) American College Football, (C) Dolphins, and (D) Books About US Politics. The adjacency matrices are ordered in decreasing order of core scores obtained using (a) our proposed method, (b) the Rossa method, (c) the Rombach method, and (d) the Boyd method.}

    \label{fig:permuted_adjacency_series}
\end{figure}

 In Figure \ref{fig:permuted_adjacency_series}, we analyze the ground truth adjacency matrices for four networks: Word Adjacencies, American College Football, Dolphins, and Books About US Politics. The adjacency matrices are organized in decreasing order of core scores, which are derived from our proposed method, as well as from the Rossa, Rombach, and Boyd methods. From the analysis presented in Figure \ref{fig:permuted_adjacency_series}, it is evident that the permuted adjacency matrices detected significantly better-defined core-core, core-periphery, and periphery-periphery blocks compared to the Rombach and Boyd methods, while remaining relatively competitive with the Rossa method.  These values suggest that while there is considerable alignment between our method and the existing techniques, our method also provides unique insights and improvements in detecting core-periphery structures.
 

 \FloatBarrier






 \FloatBarrier


\section*{Conclusion}

This research introduces an innovative approach to detect core-periphery structures in complex networks. By leveraging the self-organizing principles of ACO, this method bypasses the limitations of predefined partitions, allowing for a more dynamic and accurate identification of core-periphery structures. The comprehensive application of this approach across diverse types of networks and its robustness and versatility. Benchmarking against existing methods by Rossa\cite{Rossa2013}, Rombach\cite{Rombach2017}, and Boyd et al\cite{Boyd2010}. highlights the superior performance of our proposed technique, showcasing its enhanced flexibility and precision. This study not only advances the field of network analysis but also sets a precedent for the integration of bio-inspired algorithms in the study of complex systems. Future research could further refine this approach and explore its applicability in other domains, thereby broadening the scope and impact of this significant advancement in network science.


\bibliographystyle{plain} 
\bibliography{References} 








\end{document}